\documentclass[review]{elsarticle}

\usepackage{hyperref}
\usepackage{graphics}
\usepackage{graphicx}
\usepackage{amsfonts}
\usepackage{textcomp}
\usepackage{amssymb}
\usepackage{mathrsfs}
\usepackage{amsmath}
\usepackage{bbm}
\usepackage{color}
\usepackage{ulem}
\usepackage{adjustbox}
\usepackage{diagbox}
\usepackage{tabu}

\journal{Acta Materialia}

\begin{document}

\begin{frontmatter}

\title{Data-Driven Learning of $3$-Point Correlation Functions as Microstructure Representations}

\author[CS]{Sheng Cheng}
\author[MSE]{Yang Jiao}
\author[MAE]{Yi Ren\corref{mycorrespondingauthor}}
\address[CS]{Computer Science, Arizona State University, Tempe AZ 85287, United States}
\address[MSE]{Materials Science and Engineering, Arizona State University, Tempe AZ 85287, United States}
\address[MAE]{Aerospace and Mechanical Engineering, Arizona State University, Tempe AZ 85287, United States}

\cortext[mycorrespondingauthor]{Corresponding author: Yi Ren}
\ead{yiren@asu.edu }






\begin{abstract}
This paper considers the open challenge of identifying complete, concise, and explainable quantitative microstructure representations for disordered heterogeneous material systems. Completeness and conciseness have been achieved through existing data-driven methods, e.g., deep generative models, which, however, do not provide mathematically explainable latent representations. This study investigates representations composed of three-point correlation functions, which are a special type of spatial convolutions. We show that a variety of microstructures can be characterized by a concise subset of three-point correlations, and the identification of such subsets can be achieved by Bayesian optimization. Lastly, we show that the proposed representation can directly be used to compute material properties based on the effective medium theory.
\end{abstract}

\end{frontmatter}

\section{Introduction}


This paper considers the open challenge of identifying \textit{quantitative microstructure representations} for disordered heterogeneous material systems. 
Microstructures of such a material system can be captured as images drawn from a 2D or 3D random field.
We consider representations as the encoding of random fields in a finite-dimensional Euclidean space,  
and ideally, this encoding should be \textit{complete} and \textit{concise}.  
To explain, a representation is complete if there exists a decoder, i.e., a mapping from the representation space to the space of random fields, such that the decoded random field matches with the sample distribution of authentic microstructures used to compute the representation. 
And the representation is concise if it has minimal dimensions within all complete representations.
Both attributes are desired in a materials design context: Completeness ensures that the representation can be used for property prediction as it captures the morphological features of the material system. Conciseness enables efficient search in a low-dimensional representation space for new microstructures that potentially achieve properties beyond existing limits. 
Nonetheless, these two attributes do not guarantee explainability of predictive models built upon the representation, e.g., for structure-property prediction. And the lack of explainability hampers knowledge extraction, i.e., the natural-language understanding of how properties of a material system are affected by its morphological features.
This is what differentiates our study from existing data-driven attempts at learning microstructure representations: We seek representations that are mathematically \textit{explainable}, in addition to be complete and concise.

\begin{figure}
\centering
\includegraphics[width=0.95\textwidth,keepaspectratio]{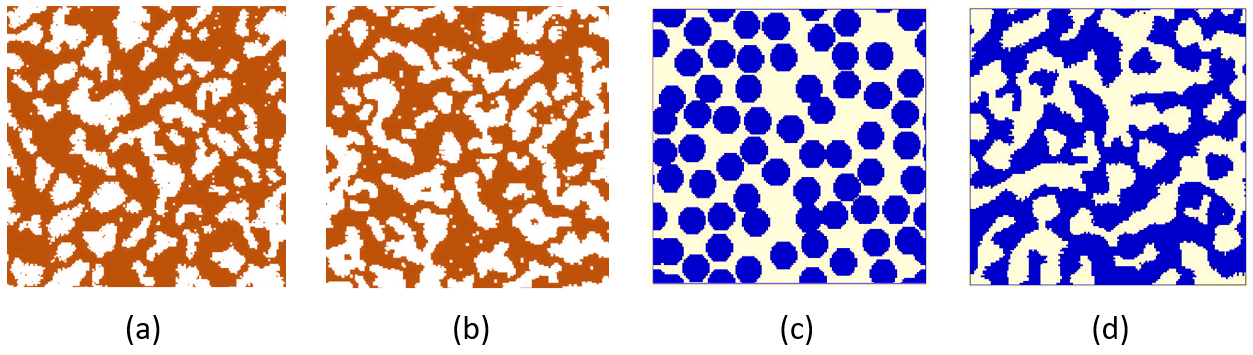}
\caption{An example of successful modeling and reconstruction of a ceramic matrix composite material [(a), original and (b), reconstruction] using the two-point correlation function $S_2$ and an example where $S_2$ is not sufficient to capture to key features of particle-reinforced composite [(c), original and (d), reconstruction].} \label{fig:introduction}
\end{figure}

To provide a background, a large body of work on representation learning exists at the intersection of materials science and machine learning. Most of these propose either complete but non-explainable, or explainable but incomplete representations. 
The former include purely data-driven generative models, e.g., restricted Boltzmann machines~\cite{cang2017microstructure}, variational autoencoders (VAE)~\cite{cang2018improving}, and generative adversarial networks (GAN)~\cite{yang2018microstructural,li2018transfer}, where a concise and near-complete representation is learned through microstructure samples, yet the encoders of which are composed of general-purpose neural networks and are non-explainable. 
The alternative approach is to use statistical models with more explainable, as opposed to learnable, architectures. These include Gaussian random fields~\cite{roberts1997statistical}, geometric descriptors (e.g., grain/particle size and shape distribution)~\cite{wilding2011clustering, callahan2012quantitative, wang2012three, ratanaphan2014five}, spectral density functions~\cite{iyer2020designing, farooq2018spectral, chen2018designing}, and $n$-point correlation functions~\cite{niezgoda2008delineation, cecen2016versatile, choudhury2016quantification, okabe2005pore, fullwood2008microstructure, jiao2007modeling, jiao2008modeling, hajizadeh2011multiple, tahmasebi2013cross, tahmasebi2012multiple, xu2014descriptor, gerke2015improving, karsanina2018hierarchical, feng2018accelerating, gao2021efficient, gao2021ultra}, among others. 
The encoding process of these methods (e.g., statistical inference and geometric characterization) are manually defined and can be explained. Yet due to the manual definitions, these representations often have limited degrees of freedom to be learned to approach completeness. To elaborate, Gaussian random fields have been used to characterize interpenetrating bi-phase morphologies (e.g., rising from the Spinodal decomposition process). This model assumes that the fluctuations in the system can be completely specified by second order statistics, which does not always hold for heterogeneous materials such as particle-reinforced composites. Geometric descriptors are typically material-specific, which are derived for microstructures composed of well-defined geometrical objects (e.g., particles/grains). 
The spectral density function is the Fourier transform of the two-point correlation function in the wave-number space, which allows one to access the large-scale correlations (via small wave-number analysis), yet have difficulty at accurately encoding small-scale morphological features that are crucial for structure-property prediction. 
Lastly, the $n$-point correlation functions encode the occurrence probabilities of specific $n$-point configurations in the microstructure~\cite{torquato1982microstructure}. 
The set of correlation functions up to infinite orders fully characterizes a random field~\cite{torquato1982microstructure, torquato2002random}, and is therefore asymptotically \textit{complete}. While it is empirically shown that some material systems can be represented by \textit{concise} sets of correlation functions, e.g., metallic alloys, ceramic matrix composites, and certain porous systems~\cite{jiao2013modeling, guo2014accurate, jiao2014modeling, chen2015dynamic, chen2016stochastic, xu2017microstructure} (Fig.~\ref{fig:introduction}a and b), there is currently a lack of systematic tools for choosing a concise and nearly complete set of correlations for any particular material system. 

Within this context, our study investigates the feasibility and efficacy of computational tools for identifying material-specific correlation functions as complete, concise, and explainable representations of disordered heterogeneous material systems. 

To elaborate on the challenge, the number of correlation functions grows exponentially as $d^{n-1}$ along the correlation order $n$ and the dimensionality $d$ of the microstructure sample (usually 2 or 3). This causes limited applications of higher-order correlation functions ($n \geq 3$) to material systems~\cite{kalidindi2011microstructure, malmir2018higher}.
On the other hand, high-order correlation functions are often required to represent disordered heterogeneous material systems~\cite{gommes2012density, gommes2012microstructural, jiao2010geometrical} (Fig.~\ref{fig:introduction}c,d). So far the choice of representative correlations relies on human intuition~\cite{chen2019hierarchical, chen2020probing}. 

As a first step towards automated selection of correlations, this paper empirically verifies the following hypothesis: {\it A concise and nearly-complete representation composed of 3-point correlations can be learned from a dataset of microstructure samples, for which 2-point correlations do not compose a complete representation.} To the authors' best knowledge, this study is the first to demonstrate the feasibility of representation learning using 3-point correlations. The following technical contributions lay the foundation for computationally tractable learning of higher-order correlations:

\begin{itemize}
    \item (i) We show that the encoder, i.e., the computation of $n$-point correlations from microstructure samples, can be modeled as a special convolutional neural network, allowing leverage of GPU computing and an existing deep learning framework.
    \item (ii) We show that the decoder, i.e., the reconstruction of microstructures, can be formulated as a topology optimization problem, allowing leverage of the state-of-the-art gradient-based algorithm which drastically improves the reconstruction efficiency from standard stochastic methods such as simulated annealing.
    \item (iii) Building on top of (i) and (ii), we show that Bayesian Optimization can effectively identify a concise subset of correlations as a nearly-complete representation. 
    \item (iv) We demonstrate that the representations resulted from (iii) can be directly utilized in combination of the effective medium theory to make accurate and explainable predictions of physical properties of heterogeneous material systems, which further supports the claim that they are nearly complete.  
\end{itemize}






The rest of the paper is organized as follows: Sec. 2 describes the models and algorithms for learning correlation functions and demonstrates using a toy case. In Sec. 3, the proposed method is applied to five material systems: porous materials (i.e., sandstone), metal-ceramic composites, metallic alloys, concrete microstructure, and particle-reinforced composites, which possess a variety of distinct structural features, including both well separated compact domains with different degrees of size and shape dispersion and interpenetrating phase morphology with varying degrees of clustering. In Sec. 4, we show that the learned representations can provide reasonable estimates of the physical properties of the material systems via the effective medium theory. In Sec. 5, we provide concluding remarks and future directions.

\section{Methods}
\label{sec:method}

To overview, our method includes three components: (i) An encoder that computes $n$-point correlations via image convolution; (ii) a decoder that performs gradient-based reconstruction of microstructures; and (iii) a search algorithm for selecting a small and nearly-complete set of $n$-point correlations. 
In the subsequent discussion, we describe each component in details, which are further illustrated through a toy case of learning 3-point correlations to represent an isosceles triangle. 

\subsection{Notations and preliminaries}
\textbf{Notations}: 
The following notations will be used throughout the paper: The indicator function $\mathbbm{1}(\textbf{c})$ returns 1 for all elements of the Boolean vector $\textbf{c}$ that are true and 0 otherwise. $\delta(x, a)=1$ if $x=a$ or otherwise 0. $\textbf{1}$ is a vector with all components equal to 1. $||\cdot||_p$ is the $l_p$ norm, and $|\cdot|$ measures the size of a countable set. $\textbf{C}(\cdot,\cdot; l, l'): \mathcal{A} \times \mathcal{B} \rightarrow \mathbb{R}^{l\times l}$ denotes the image convolution that discretizes the input $a \in \mathcal{A}$ and the filter $b \in \mathcal{B}$ as $l$-by-$l$ and $l'$-by-$l'$ matrices, respectively.

For ease of exposition, we will focus on 2D bi-phase microstructures, although the presented method can be extended to 3D multi-phase microstructures. A microstructure sample is an indicator function $y: \mathcal{X} \rightarrow \{0,1\}$ where $\mathcal{X} = [0,L]^d$ is the support and $L$ is its linear size. 
We denote by $\mathcal{Y}$ the space of $y$, and $\tilde{\mathcal{Y}}=\{y_i\}_{i=1}^N \subset \mathcal{Y}$ a finite dataset that contains $N$ microstructure samples of the same material system. The empirical data distribution is denoted by $p_{\tilde{\mathcal{Y}}}$.

\begin{figure}
    \centering
    \includegraphics[width=0.5\textwidth]{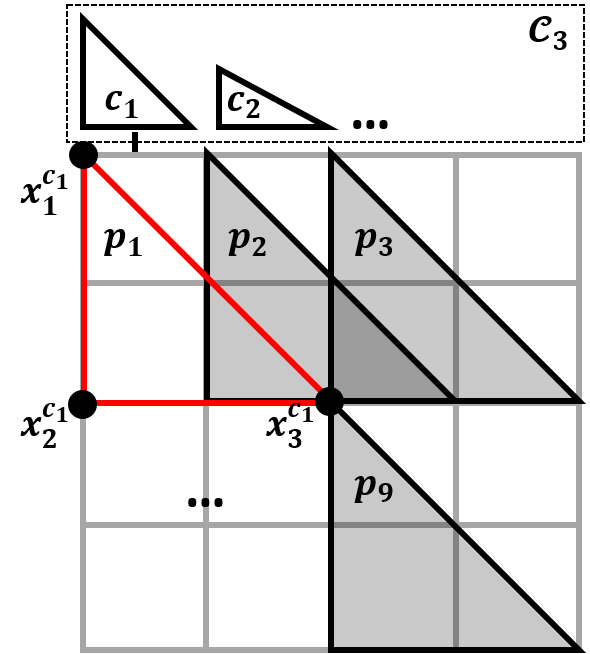}
    \caption{Visualization of the 3-point correlation set $\mathcal{C}_3=\{c_1, c_2,...\}$, the polytopes $\{p_1,p_2,...p_9\}$ of $c_1$, and the configuration-specific vertices $\{{\bf x}_1^{c_1},...,{\bf x}_3^{c_1}\}$}
    \label{fig:correlation}
\end{figure}

\textbf{$n$-point correlation functions}: Given $n$, an $n$-point configuration space $\mathcal{C}_n \in \mathbb{R}^m$ can be constructed, where $m= d^{n-1}$ is the degrees of freedom of the $n$-point configuration. To explain using 2D examples ($d=2$): When $n = 2$, $\mathcal{C}_2 = [0,\infty) \times [0,\pi)$ contains all distance-angle pairs that define line segments in $\mathbb{R}^2$; when $n = 3$, $\mathcal{C}_3 = [0,\infty)^{3}\times [0,\pi)$ contains all distances-angle tuples that define triangles. Note that $\mathcal{C}_n$ is a subspace of $\mathcal{C}_{n+1}$. Therefore $\mathcal{C}_{\infty}$ contains all correlation configurations. Each correlation configuration $c \in \mathcal{C}_{\infty}$ defines a set of polytopes $\mathcal{P}_c$ and each polytope $p \in \mathcal{P}_c$ is defined by a set of vertices $\{{\bf x}_1^p,...,{\bf x}_n^p\}$ translated from a configuration-specific set $\{{\bf x}_1^c,...,{\bf x}_n^c\}$ (Fig.~\ref{fig:correlation}). Given $y$, an $n$-point correlation $s: \mathcal{C}_{\infty} \times \mathcal{Y}  \rightarrow [0,1]$ for the phase of interest (assumed to be $y(\cdot)=1$) is defined as
\begin{equation}
    s(c,y) = \frac{1}{|\mathcal{P}_c|}\sum_{\{{\bf x}_1^p,...,{\bf x}_n^p\} \in  \mathcal{P}_c} \mathbbm{1}\left(y({\bf x}_1^p)=y({\bf x}_2^p)=...=y({\bf x}_n^p) = 1\right),
\end{equation}
which is the probability that a uniformly selected polytope $p \in \mathcal{P}_c$ has vertices all belonging to the phase of interest~\cite{torquato2002random}. The order of $s$ is determined by $c$.


\subsection{Computing $n$-point correlations through image convolution}
We now show that $n$-point correlations can be computed through image convolution, and therefore a standard convolutional neural network architecture can be adopted. Let convolution kernels parameterized by an $n$-point configuration $c$ be a function $w_c: \mathcal{X} \rightarrow \{0,1\}$:
\begin{equation}
    w_c(x) = \sum_{x_i \in \{{\bf x}_1^c,...,{\bf x}_n^c\}} \delta(x, x_i).
\end{equation}
We can now compute the correlation as
\begin{equation}
    s(c,y) = \frac{ \textbf{1}^T \mathbbm{1}\left(\textbf{C}(y,w_c;l,l') \geq n \right)\textbf{1}}{l^2}.
\end{equation}
Since the indicator function is not differentiable, we propose the following approximation:
\begin{equation}
    \hat{s}(c,y) = \frac{ \textbf{1}^T \left(\sigma(\textbf{C}(y,w_c;l,l')-n+1)\right)\textbf{1}}{l^2},
\end{equation}
where $\sigma(x) = x$ if $x>0$ and 0 otherwise, and has differentiable implementations~\cite{paszke2017automatic}. Note that for bi-phase microstructures, the approximation is exact. 
Lastly, we denote by 
$\hat{s}(c,\tilde{\mathcal{Y}}) = \mathbb{E}_{Y \sim p_{\tilde{\mathcal{Y}}}} \left[\hat{s}(c,Y)\right]$ the mean correlation of the dataset with respect to $c$.

\subsection{Gradient-based microstructure reconstruction}
Given a set of correlations $\hat{\mathcal{C}} \in \mathcal{C}_{\infty}$, a dataset $\tilde{\mathcal{Y}}$, and target correlations $\{\hat{s}(c,\tilde{\mathcal{Y}})\}_{c\in \hat{\mathcal{C}}}$, the reconstruction problem is to find a microstructure $y$ with correlations close to the target. With discretization of $\mathcal{X}$ and a focus on bi-phase microstructures, $y$ can be considered as a $l$-by-$l$ binary matrix. This leads to the following topology optimization problem:
\begin{equation}
    \min_{y \in \{0,1\}^{l^2}} \quad L_{\hat{\mathcal{C}}}(y) := \sum_{c\in \hat{\mathcal{C}}} \left(\hat{s}(c,y) - \hat{s}(c,\tilde{\mathcal{Y}})\right)^2.
    \label{eq:recon}
\end{equation}

We solve Eq.~\eqref{eq:recon} using gradient descent with the following techniques. \textbf{Relaxation to a continuous problem:} Note that Eq.~\eqref{eq:recon} is combinatorial when $y$ is binary. Therefore it is standard to solve a relaxed problem by considering $y$ as a soft step function of another variable $u$: 
\begin{equation}
    y(u;\beta) = \frac{\tanh(\beta / 2) + \tanh(\beta (u - 1/ 2))}{2\tanh(\beta / 2)},
\end{equation}
where $u\in \mathbb{R}^{l \times l}$.
We can then perform gradient-based search using the gradient $\nabla_u L_{\hat{\mathcal{C}}}(y(u;\beta)) = \nabla_y L_{\hat{\mathcal{C}}}(y) \nabla_u y(u;\beta)$. \textbf{Avoiding gradient vanishing:} The latter part of the gradient, $\nabla_u y(u;\beta)$, vanishes quickly as $u$ moves away from 0 and this issue exacerbates with larger $\beta$. On the other hand, large $\beta$ is necessary for $y$ to approach a binary solution. A plausible solution to this dilemma is to gradually increase $\beta$ during the search~\cite{wu2017infill}. 
Our experiments suggest that this scheme for changing $\beta$ is data dependent (see Sec.~\ref{sec:result}).
\textbf{Removing artifacts in reconstructions:} When solving topology optimization problems, gradient descent can lead to local and artificial checkerboard patterns in the solution. To address this issue, we apply a Gaussian filter $w_g$ to the gradient at every search step, where $w_g$ is a $3$-by-$3$ Gaussian kernel with a mean of 0 and a standard deviation of 3. The standard deviation is set based on experiments.

\subsection{A goodness metric for choosing correlations}

Let $p_{\hat{\mathcal{C}}}$ be the distribution of reconstructed microstructures resulted from Eq.~\eqref{eq:recon} given a correlation set $\hat{\mathcal{C}}$. The randomness is due to the random initialization of $u$. Here we discuss the procedure for choosing a concise set $\hat{\mathcal{C}}$ so that $p_{\hat{\mathcal{C}}}$ is representative of $p_{\mathcal{Y}}$. 
To start, we propose the following problem:
\begin{equation}
    \min_{\hat{\mathcal{C}} \subset \mathcal{C}} \quad L(\hat{\mathcal{C}}) = \phi(p_{\hat{\mathcal{C}}}, p_{\tilde{\mathcal{Y}}}),
    \label{eq:bo}
\end{equation}
where $\phi(\cdot, \cdot)$ measures a distance between two distributions. We note that morphological consistency between $p_{\hat{\mathcal{C}}}$ and $p_{\tilde{\mathcal{Y}}}$ is of higher priority than matching between the distributions, i.e., we consider $\hat{\mathcal{C}}$ a good representation as long as samples from $p_{\hat{\mathcal{C}}}$ are morphologically similar to $\tilde{\mathcal{Y}}$, even when the support of $p_{\hat{\mathcal{C}}}$ does not fully overlap with that of $p_{\tilde{\mathcal{Y}}}$. With this insight, we choose to define $\phi$ on the full correlation set $\mathcal{C}_{\infty}$:
\begin{equation}
    \phi(p_1, p_2) = \mathbb{E}_{y_1 \sim p_1, y_2 \sim p_2} \left[ \frac{1}{|\mathcal{C}_{\infty}|}\sum_{c \in  \mathcal{C}_{\infty}} (\hat{s}(c,y_1) - \hat{s}(c,y_2))^2 \right].
    \label{eq:theoretical_distance}
\end{equation}
However, Eq.~\eqref{eq:theoretical_distance} is ill-defined since $\mathcal{C}_{\infty}$ is an infinite-dimensional space, and approximating $\mathcal{C}_{\infty}$ with large $n$ is intractable. In fact, if we discretize each dimension of $\mathcal{C}_n$ by $m$ levels, there will be $m^{d^{(n-1)}}$ correlation values to compute for every $y$. As practical solutions, in this paper we will introduce task-dependent and computationally tractable surrogates (denoted by $\hat{\phi}$)) as our best effort to approximating $\mathcal{C}_{\infty}$ (see Sec.~\ref{sec:result} for details).

\subsection{Search of representations}
Since computing $p_{\hat{\mathcal{C}}}$ relies on the reconstruction, which is challenging to be differentiated, we propose to solve Eq.~\eqref{eq:bo} using grid search and Bayesian optimization (BO). For both algorithms, we fix the subset size $|\hat{\mathcal{C}}|$. Grid search enumerates over all combinations of correlations using a discretized $\mathcal{C}_n$; BO actively samples such combinations without full enumeration. 
Both algorithms suffer from the curse of dimensionality. To alleviate this challenge in the material cases in Sec.~\ref{sec:result}, we define $\hat{\mathcal{C}}$ as drawn from a distribution on $\mathcal{C}_{\infty}$, and search only for the low-dimensional parameters of this distribution.

\subsection{A toy case}
\begin{figure}
    \centering
    \includegraphics[width=\textwidth]{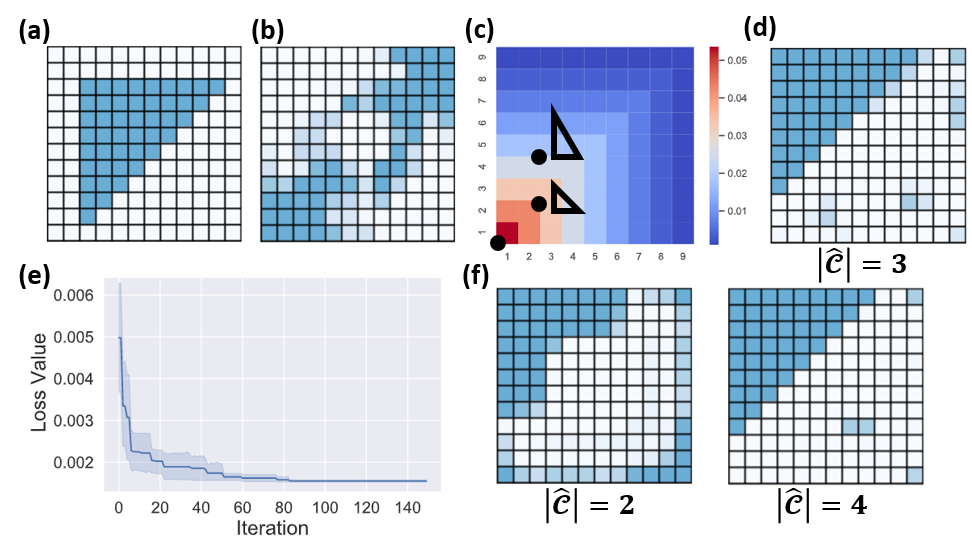}
    \caption{Bayesian optimization for finding a concise representation of a filled isosceles right triangle in (a) using 3-point correlations. (b) Reconstruction fails with a full set of 2-point correlations. (c) Bayesian optimized representations visualized as triangles (along with the volume fraction at the origin) in the 3-point correlation landscape spanned by the polytope parameters. (d) Resultant reconstruction based on (c). (e) BO convergence. (f) Bayesian optimized reconstructions with two and four correlations.}
    \label{fig:toycase}
\end{figure}

Here we use a toy case to walk through the algorithmic settings for the encoding, decoding, and Bayesian optimization. It should be noted that same details, including the parameterization of $\hat{\mathcal{C}}$ and the choice of $\hat{\phi}$ are case-dependent.
\textbf{Case setup}: Let the dataset be 2D binary images, where each image $y \in \{0,1\}^{12 \times 12}$ contains a filled isosceles right triangle. The goal is to find a subset of correlations that reconstruct the triangles. Since correlations are invariant to linear translation, it is sufficient to keep a single image as the dataset (Fig.~\ref{fig:toycase}a). While primitive, this case demonstrates the need for higher-order correlations for reconstruction: Fig.~\ref{fig:toycase}b shows the inferior reconstruction result using the full set of 2-point correlations. 
\textbf{Reconstruction}: We initialize a deterministic reconstruction process by setting the initial guess to $y(\cdot)=0$, and increase $\beta$ from 7 to 15 with an interval of 2. $\beta$ is increased for every 2000 iterations of gradient descent on the relaxed version of Eq.~\eqref{eq:recon}. The step size for gradient descent is set to 2000 for $\beta \leq 11$ and 1000 for larger $\beta$.
\textbf{Parameterization of the correlation space}: For the toy case, we choose to use a subset of three 3-point correlations for this toy study: $\hat{\mathcal{C}}=\{c_1,c_2,c_3\}$, with the following parameterization. First, we set $c_1$ as the volume fraction (i.e., with all three vertices overlap). For $c_2$ and $c_3$, we fix one vertex and allow the other two to move in the horizontal and vertical directions, respectively, within a 18-by-18 box. This yields a 4-dimensional continuous and bounded space where $\hat{\mathcal{C}}$ will be searched from. \textbf{Criterion for choosing correlations}: For the toy case, we replace $\mathcal{C}_{\infty}$ with $\mathcal{C}_3$ in Eq.~\eqref{eq:theoretical_distance}, based on the empirical evidence to be presented that 3-point correlations are sufficient for reconstructing the triangle. 
\textbf{Bayesian optimization}: We start the search with 100 samples from the four dimensional search space where the samples are drawn using a Latin Hypercube sampler. Each sample defines a correlation set from which $\hat{\phi}$ can be computed. With this data, a Gaussian process model can be built on the search space, which allows prediction of $\hat{\phi}$ values and their uncertainties for any correlation set. With this, we search for a new correlation set that has the maximum expected improvement in $\hat{\phi}$ from the current samples. Once sampled, this new data point is added to the existing dataset to update the Gaussian process model. This process continues for 150 iterations.


\textbf{Results and remarks}:  
Fig.~\ref{fig:toycase}b-e summarizes the resultant $\hat{\mathcal{C}}$, the corresponding reconstruction, and the BO loss history. Due to the probabilistic nature of the algorithm, loss variances are estimated over five independent BO trials. The results suggest that a subset of three 3-point correlations is nearly complete for reconstructing the given sample. To show that the resultant representation is concise, we compare the reconstruction results from using two, three, and four 3-point correlations in Fig.~\ref{fig:toycase}d,f. The correlations are chosen by BO in all cases. 
This toy case demonstrates that a concise subset of 3-point correlations can be used to reconstruct images. In the next section, we apply this idea to material microstructures. 

\section{Experimental Results}
\label{sec:result}

We test five material systems, including sandstone, particle-reinforced composites, concrete micro-structure, metal-ceramic composites and metallic alloys. These material systems possess a variety of distinct structural features, including both well separated compact domains with different degrees of size and shape dispersion and interpenetrating phase morphology with varying degrees of clustering. 
Our experiments suggest that a rich variety of morphologically consistent microstructure samples can be reconstructed using the proposed method, even when a single microstructure sample is used for computing the representations. Therefore, all experiments will be based on a single microstructure sample from each of the material systems. We will focus on sandstone to elaborate on the algorithmic settings.     


\begin{figure}
    \centering
    \includegraphics[width=0.3\textwidth]{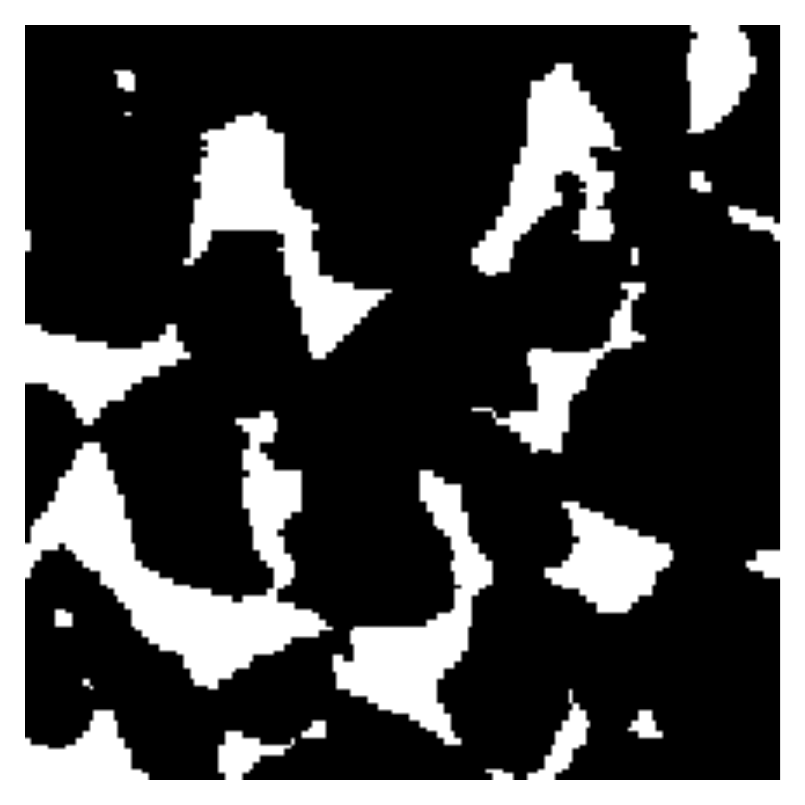}
    \caption{Sandstone sample}
    \label{fig:sandstone}
\end{figure}

\textbf{Case setup}:
A sandstone microstructure sample $y \in \{0,1\}^{128 \times 128}$ is composed of the rock (shown in black) and the pore phase (shown in white). Fig.~\ref{fig:sandstone} shows 2D slices of a 3D sandstone sample obtained via X-ray tomography~\cite{li2018accurate}. It can be seen that the sedimented rock grains possess a wide shape and size dispersion, leading to complex morphology of the pore space. The material possesses a statistically homogeneous and isotropic microstructure so that 3-point correlation information extracted from the 2D images are also representative of the full 3D system~\cite{malmir2018higher}. 

\textbf{Reconstruction}: We initialize with $\beta = 5$ and double its value up to 20 after a fixed amount of gradient descent steps. This amount is set to 50k for
$\beta \leq 13$ and 10k otherwise. The step size for each step is set to 2000. The initial guess is randomized by sampling each pixel value from a Bernoulli distribution parameterized by the volume fraction of the given microstructure.

\textbf{Parameterization of the correlation space}: For each correlation, we fix one vertex at the center and allow the other two to move in a 41-by-41 box with a stride of 2. To define a search space of correlations, we hypothesize that both the number and the length scales of correlations affect the reconstruction quality. Specifically, given a maximum length scale, more correlations leads to better reconstruction, yet we expect that the benefit of adding more correlations diminishes. On the other hand, for a fixed number of correlations, there exists an optimal maximum length scale within which the correlations are representative. With these hypotheses, we propose to search on a 2D grid with the maximum correlation length scale as $l \in \{13, 21, 29, 37\}$ and the number of correlations as $k \in \{\frac{1}{64}K, \frac{1}{32}K, \frac{1}{16}K, \frac{1}{10}K, \frac{1}{8}K\}$ where $K=7381$ is the total number of correlations when the maximum length scale is 21. These parameter ranges are empirically chosen.  Experimental results suggest that larger $l$s and $k$s contribute minimally to reconstruction improvements. Each grid point $(k, l)$ defines a distribution of representations, each of which is composed of $k$ 3-point correlations which are uniformly sampled from the set of correlations with a maximum length scale of $l$.  

\textbf{Criterion for choosing correlations:} From experimental results, we propose to evaluate the reconstruction results using the {\it pore-size distribution function} $P(r;y)$ as a surrogate. Specifically, $P(r;y)dr$ provides the probability that a spherical ``cavity'' of radius $r$ can be entirely inserted into the phase of interest centered at a randomly selected point in that phase of $y$~\cite{torquato2002random}. $P(r;y)$ was originally introduced to quantify the void phase in porous materials and was subsequently generalized as a generic statistical descriptor to quantify disordered heterogeneous materials. Since the porous phase is hosting all transport processes such as fluid flow and chemical diffusion in the porous media, the function $P(r;y)$ is shown to quantitatively related to a variety of transport properties including effective diffusivity, fluid permeability and mean-survival time of chemicals~\cite{torquato2002random}. Moreover, $P(r;y)$ essentially provides a spherical measure of the clustering information in the phase of interest and thus encodes information on higher-order correlations~\cite{jiao2009superior}. In this study, we set $r \in \{0,..., 63\}$ pixels and define
\begin{equation}
    \hat{\phi}(p_1,p_2) = \mathbb{E}_{y_1 \sim p_1, y_2 \sim p_2}\left[\frac{1}{64} \sum_{r \in \{0, ..., 63\}}\left | P(r;y_1)-P(r;y_2)\right | \right].
    \label{eq:reconstruction_p}
\end{equation}

\begin{figure}
    \centering
    \includegraphics[width=1.0\textwidth]{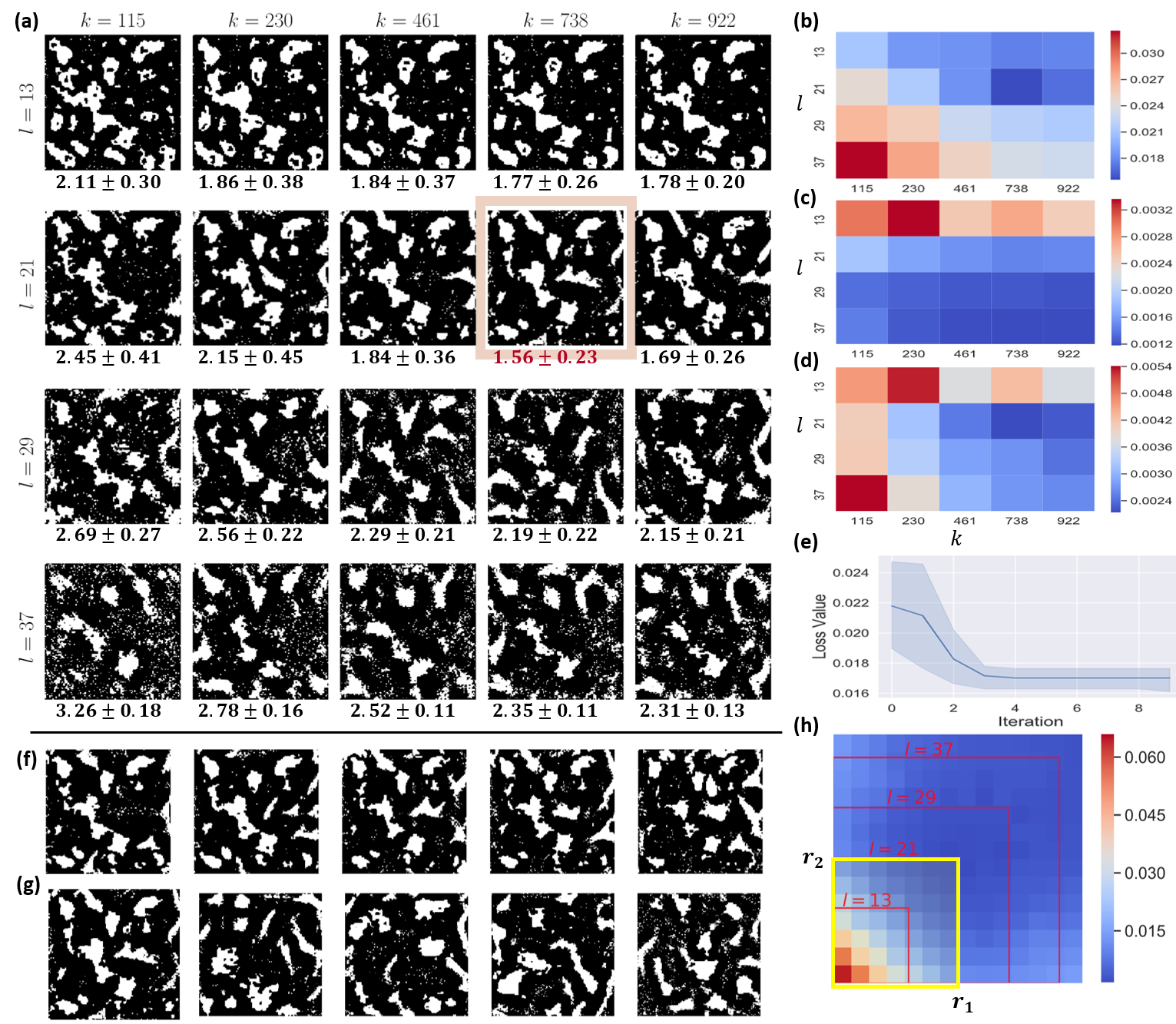}
    \caption{(a) Reconstructions of sandstone with different numbers of correlations $k$ and maximum correlation length scales. Sample means and standard deviations of pore-size distribution function values are reported based on 25 randomly drawn correlation sets from the corresponding $(k,l)$. Unit: $\times 10^{-2}$.  (b-d) The pore-size distribution, $l_1$ loss, and weighted $l_1$ loss on the $(k,l)$ grid. (e) Convergence of Bayesian optimization. (f, g) Reconstructions by random initial guesses and random correlation sets drawn from the optimal $(k,l)$, respectively. (h) Correlation landscape and the optimal maximum length scale $l$.}
    \label{fig:design}
\end{figure}

\textbf{Bayesian Optimization}: Since we have a relatively small candidate set, BO is started with five configurations of $(k,l)$. For each configuration, we compute $\hat{\phi}$ following Eq.~\ref{eq:reconstruction_p} with five draws of correlations from the uniform distribution specified by $(k,l)$. BO terminates with nine samples. See Fig.~\ref{fig:design}e for BO convergence, where loss variances are estimated over five independent BO trials. A full grid search is performed to verify the efficacy of BO. See Fig.~\ref{fig:design}a for reconstruction results and pore-size values from sampled correlations for all $(k,l)$ configurations and the same initial guess. Standard deviations are estimated from five correlation sets drawn independently based on the corresponding $(k,l)$. The Bayesian optimized configuration is consistent with the grid search. 



\textbf{Remarks}: \textit{(1) How much is the variance in reconstruction quality due to the random initialization and the random sampling of correlations?} 
We draw 25 initial guesses of $u$ from independent Bernoulli distributions parameterized by the volume fraction of the given sandstone sample. The sample variance in reconstruction quality is 3.31e-3.
Fig.~\ref{fig:design}f shows five sample reconstructions.
 Since the variance is small, we do not use multiple random initialization for reconstruction during BO.

Fig.~\ref{fig:design}a reports variances in reconstruction quality for all $(k,l)$ configurations due to random sampling of correlations. Variances are estimated based on 25 random draws of correlations for each grid point using the same initialization of $u$. While the variances are not negligible, they are small enough so that an optimal configuration ($k=738$ and $l=21$) can be identified with statistical significance. We demonstrate reconstruction from five randomly drawn correlation sets for the optimal configuration in Fig.~\ref{fig:design}g.


\textit{(2) Is pore-size distribution a good surrogate of $\phi$?} 
Visual comparison in Fig.~\ref{fig:design}a supports the use of pore-size distribution: The optimal configuration achieves reconstruction quality no worse than others.
We provide a second evidence by comparing the pore-size results with those from alternatives that are based on the full set of 3-point correlations with a maximum length scale of 41. Minimal variations in correlations are observed beyond this length scale. With minor abuse of notation, we will denote this set by $\mathcal{C}_3$, and 3-point correlation vector of an input $y$ by $c(y) = [c_1(y),...,c_{|\mathcal{C}_3|}(y)]$, where $c_i(y)$ is the correlation of $y$ according to the $i$th element of $\mathcal{C}_3$. The alternatives are based on the $l_1$ norm (Fig.~\ref{fig:design}c) and the weighted $l_1$ norm of $c(y)-c(y')$, where $y$ and $y'$ are the microstructure sample and its reconstruction, respectively. We notice that the sizes of subsets of $c_i(y)$ corresponding to different length scales increases along the length scale, i.e., there are more correlations evaluated on large triangles than small ones in $c(y)$. Since correlations asymptotically approach zero as the length scale increases, $c(y)$ will increasingly be dominated by zero elements when the maximum length scale increases, making reconstruction qualities less distinguishable between two choices of $k$ when using $l_1$ norm as a metric (See Fig.~\ref{fig:design}c). To address this, we introduce a weighted norm as $\sum_{i=1}^{|\mathcal{C}_3|} w_i |c_i(y)-c_i(y')|$, where $w_i$ is the proportion of the subset of correlations of the length scale where $c_i$ belongs. 
This leads to an optimal representation consistent with that from using the pore-size distribution function (Fig.~\ref{fig:design}d). However, the choice of reweighting is not grounded in theory and can potentially be case dependent. 

\textit{(3) Is the chosen representation explainable?} From the grid search results, we conjecture that there is a critical length scale within which correlations are representative, because when we choose larger $l$s, some of the randomly drawn correlations will have larger length scales that are not representative, thus lowering the reconstruction quality (see rows for $l=29$ and $37$ in Fig.~\ref{fig:design}a,b). To better understand this, we examine the 3-point correlation landscape over a range of length scales, as shown in Fig.~\ref{fig:design}h. For visualization purpose, we parameterize the three points of a polytope $p$ using $r_1 = ||\textbf{x}_1^p-\textbf{x}_2^p||_2$, $r_2 = ||\textbf{x}_1^p-\textbf{x}_3^p||_2$, and set the angle $\theta=0 \deg$ at vertex $x_1$. Each grid point in Fig.~\ref{fig:design}h represents the average correlation value over all orientations of the polytope defined by $r_1$ and $r_2$. We see that all variations in correlations occur within a length scale, which corresponds well to the optimized one. The same characteristic length scale can also been seen on correlation landscapes with $\theta=45 \deg$ and $\theta=90 \deg$. 

\begin{figure}
    \centering
    \includegraphics[width=1.0\textwidth]{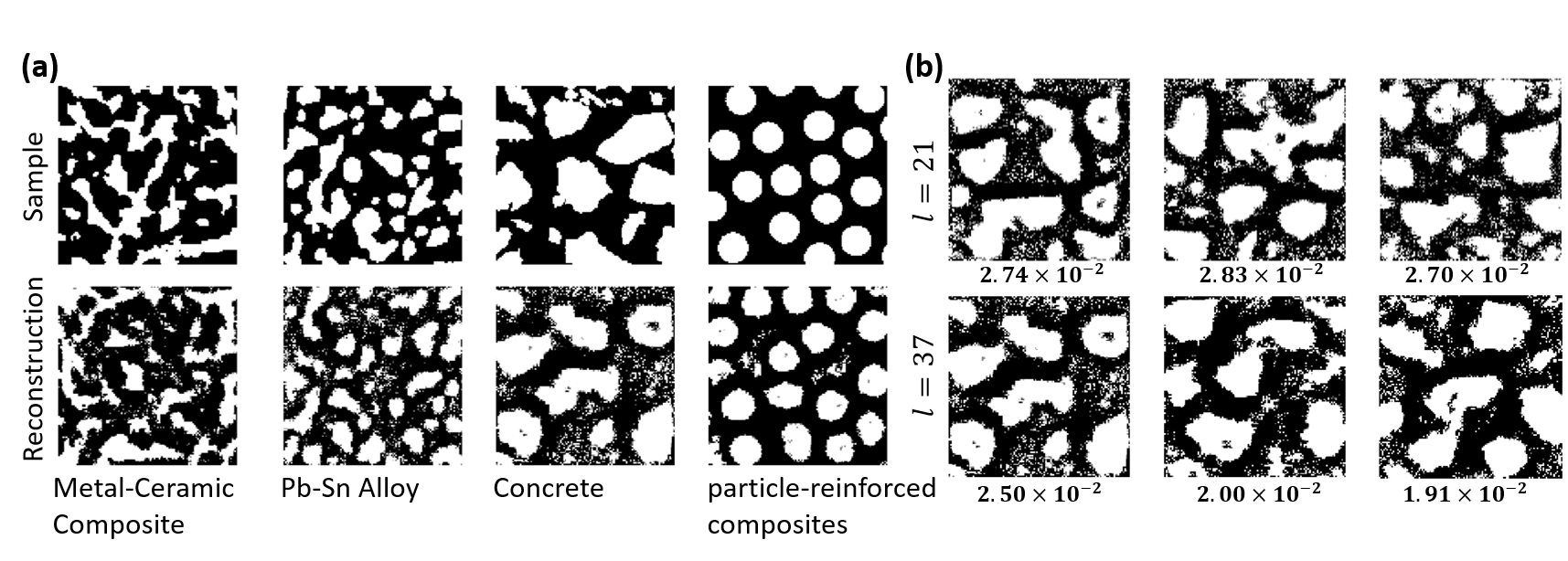}
    \caption{(a) Reconstructions of new materials using the optimal $(k,l)$ from sandstone (b) Increasing the maximum length scale for concrete improves its reconstruction quality (measured by the pore-size distribution function)}
    \label{fig:more}
\end{figure}

\textit{(4) Can the chosen representation be generalized to other material systems?} 
We now test whether the learned representation for sandstone can be applied to other binary-phase material systems.
From (3), we hypothesize that the representation may fail on a new material system if its microstructure has a different characteristic length scale from sandstone. 
Results on metal-ceramic composite, Pb-Sn alloy, concrete, and particle-reinforced composite are summarized in Fig.~\ref{fig:more}a. The reconstruction quality is acceptable for all material systems except for concrete. We conjecture that this is because metal-ceramic composite, Pb-Sn alloy, and particle-reinforced composite possess similar morphology features as those for sandstone, e.g., distinct local (short-range) correlations mainly determined by the shape and size of local structural elements (e.g., particles) and their short-range ordering, which then quickly decay to the average asymptotic values.
Concrete, on the other hand, has larger characteristic length scale than that of sandstone. This hypothesis is empirically tested in Fig.~\ref{fig:more}b where we show that increasing $l$ improves the reconstruction quality of concrete.

\section{Property Predictions via Optimized Correlations}
\label{sec:property}

In this section, we show that the optimized representations can directly provide reasonable estimates of the physical properties of the material systems. This is achieved by employing the effective medium theory, in particular, the so-called {\it strong contrast expansion} (SCE) formalism which allows one to analytically express material properties such as elastic modulus and thermal/electrical conductivity as an infinite series of integrals of $n$-point correlation functions convoluted with the proper field kernels~\cite{torquato1997effective, torquato1997exact}, given the properties of individual material phases. It has been shown that the SCE formalism possess superior convergence behavior such that truncation of the infinite series at relatively lower order $n$ can already yield very accurate estimates of the physical properties of interest for many heterogeneous material systems~\cite{li2016microstructural}.


Here, we focus on mechanical properties, including the overall (or effective) bulk modulus $K_e$ and the shear modulus $G_e$, which are well defined for all of the material systems investigated here. In particular, the 3-point SCE approximation obtained by truncating the SCE series at $n = 3$ for the effective bulk modulus $K_e$ is given by
\begin{equation}
\label{bulk} \phi_2 \frac{\kappa_{21}}{\kappa_{e1}} = 1 -
\frac{(d+2)(d-1)G_1\kappa_{21}\mu_{21}}{d(K_1 + 2G_1)} \phi_1 \chi
\end{equation}
Similarly, the 3-point SCE approximation obtained by truncating the SCE series at the $n = 3$ for the effective shear modulus $G_e$ is given by
\begin{equation}
\label{shear} \phi_2 \frac{\mu_{21}}{\mu_{e1}} = 1-
\frac{2G_1\kappa_{21}\mu_{21}}{d(K_1 + G_1)}\phi_1 \chi -
\frac{2G_1\kappa_{21}\mu_{21}}{d(K_1 + 2G_1)}\phi_1 \chi -
\frac{1}{2d}[\frac{dK_1 + (d-2)G_1}{K_1 + 2G_1}]^2\mu_{21}^2\phi_1
\eta_2
\end{equation}
where $\phi_1$ and $\phi_2$ are respectively the volume fraction of
the black (``matrix'') and the white (``inclusion'') phases; $d$ is the spatial dimension of the material system; $K_p$ and $G_p$ are respectively
the bulk and shear modulus of phase $p$; the scalar parameters
$\kappa_{pq}$ and $\mu_{pq}$ ($p, q = 1, 2, e$) are respectively
the bulk and shear modulus polarizability, defined as
\begin{equation}
\kappa_{pq} = \frac{K_p - K_q}{K_q + \frac{2(d-1)}{d}G_q}
\end{equation}
and
\begin{equation}
\mu_{pq} = \frac{G_p - G_q}{G_q}\frac{1}{1+\frac{\frac{d}{2}K_q +
\frac{(d+1)(d-2)}{d}G_1}{K_q + 2G_q}};
\end{equation}
and $\chi$ and $\eta_2$ are the microstructural
parameters associated with the phases of interest incorporating the
3-pt correlations $s_3$ as well as 2-pt correlation correlations $s_2$ (which are certain subset of $s_3$), i.e.,
\begin{equation}
\label{chi} \chi=
\frac{9}{2\phi_1\phi_2}\int_0^{\infty}\frac{dr}{r}\int_0^{\infty}\frac{ds}{s}\int_{-1}^{1}d(\cos\theta)P_2(\cos\theta)[s_3(r,
s, t)-\frac{s_2(r)s_2(t)}{\phi_2}]
\end{equation}
and
\begin{equation}
\eta_2 = \frac{5}{21}\chi +
\frac{150}{7\phi_1\phi_2}\int_0^{\infty}\frac{dr}{r}\int_0^{\infty}\frac{ds}{s}\int_{-1}^{1}d(\cos\theta)P_4(\cos\theta)[s_3(r,
s, t)-\frac{s_2(r)s_2(t)}{\phi_2}],
\end{equation}
where $t = (r^2 + s^2 - 2 rs \cos\theta)^{1/2}$, and $P_2$ and
$P_4$ are respectively the Legendre polynomials of order two and
four, i.e.,
\begin{equation}
P_2(x) = \frac{1}{2}(3x^2 - 1), \quad P_4(x) =
\frac{1}{8}(35x^4-30x^2+3).
\end{equation}

\begin{table}[htp]
\centering 
\begin{tabular}{c|@{\hspace{0.25cm}}c@{\hspace{0.25cm}}c@{\hspace{0.25cm}}c@{\hspace{0.25cm}}c}
\hline\hline
Materials  & $K_1$ & $G_1$ & $K_2$ & $G_2$\\
\hline
Sandstone (rock \& void)      & 35.8 & 10.2 & 0 & 0 \\\hline
Metal-ceramic Composite (SiC \& Al)     &  180 & 110 & 68 & 24 \\
Pb-Sn Alloy    &   42 & 14 & 45.8 & 4.9  \\\hline
Concrete (cement \& rock)    &  32 & 9 & 38 & 11.5 \\\hline
Particle-reinforced Composite (Al \& SiC)  &   68 & 24 & 180 & 110 \\\hline
\hline\hline
\end{tabular}
\caption{Individual phase properties ($K_i$ and $G_i$) for the heterogeneous material systems investigated in this work. The unit for the bulk and shear modulus are GPa.}
\label{tab1}
\end{table}


\begin{table}[htp]
\centering 
\begin{tabular}{c|@{\hspace{0.25cm}}c@{\hspace{0.25cm}}c@{\hspace{0.25cm}}c@{\hspace{0.25cm}}c|@{\hspace{0.25cm}}c@{\hspace{0.25cm}}c@{\hspace{0.25cm}}c@{\hspace{0.25cm}}c}
\hline\hline
Materials & $K_L$ & $\bar{K}_e$ & $K_e$ & $K_U$ & $G_L$ & $\bar{G}_e$ & $G_e$ & $G_U$\\
\hline
Sandstone       & 0 & 16.7$\pm$0.9   & 17.3 & 19.3 & 0 & 5.1$\pm$0.8 & 5.5 & 7.1 \\\hline
Metal-ceramic Composite     & 132.2 & 134$\pm$3.8   & 136.9 & 141.0 & 68.4 & 70$\pm$1.7 & 74.3 & 78.8  \\\hline
Pb-Sn Alloy   & 43 & 41.5$\pm$0.8   & 43.02 & 43.1 & 10.3 & 9.6$\pm$0.7 & 10.4 & 10.8   \\\hline
Concrete   & 33.8 & 31.4$\pm$0.7   & 34.0 & 34.4 & 9.8 & 9.1$\pm$0.4 & 9.9 & 10.2  \\\hline
Particle-reinforced Composite  & 86.9 & 87$\pm$2.9   & 89.2 & 96.3 & 37.9 & 38$\pm$1.5 & 38.8 & 44.7  \\\hline
\hline
\end{tabular}
\caption{Comparison of the estimated elastic properties ($\bar{K}_e$ and $\bar{G}_e$) for the heterogeneous material systems using the reduced set of 3-point representations to the corresponding Hashin-Shtrikman bounds ($K_U$, $K_L$ and $G_U$, $G_L$), as well as the ground truth values ($K_e$ and $G_e$) computed from SCE using the complete set of 3-point correlations. The unit for the bulk and shear modulus are GPa.}
\label{tab2}
\end{table}

The elastic modulus $K_e$ and $G_e$ for different material systems are respectively estimated using Eq.~\eqref{bulk}) and Eq.~\eqref{shear} by computing the 3-point microstructural parameters $\chi$ and $\eta_2$ using the optimized distribution of the 3-point representations obtained from the previous section. The properties of individual phases for each material system are given in Tab.~\ref{tab1}. For each material system, 5 sets of 3-point statistics are independently sampled based on the optimized distribution, and used to compute the estimated properties using the corresponding SCE. The 2-point correlations are computed by using the same set of distances defined by edge length of the sampled triangular kernels. The averaged $K_e$ and $G_e$ are give in Tab.~\ref{tab2}, which are compared to Hashin-Shtrikman bounds~\cite{hashin1963variational}, as well as the ``ground truth'' values, which are computed from SCE using the complete set of 3-point functions. It can be seen from the table that the estimated elastic properties using the reduced set of 3-point representations are within the theoretical bounds and can provide reasonable estimates of the actual material properties.



We note again that purely data-driven models (e.g., representation via GAN or VAE and representation-property mapping via supervised learning) can also be used for property prediction~\cite{cang2018improving}. However, such methods do not provide explainable representations and require extra compute for supervised learning.

\section{Conclusion and Discussions}
\label{sec:conclusion}

We introduced a method to learn concise, complete, and explainable representations of complex heterogeneous material systems based on 3-point correlations. The key components of our method include a convolutional network architecture for efficient computation of 3-point correlations, and an algorithm for gradient-based microstructure reconstruction. Bayesian optimization or grid search is subsequently applied to obtain the optimal subset or distribution of 3-point correlations based on physics-inspired reconstruction quality metrics. 
The utility of our method is demonstrated in detail via the quantification, modeling and reconstruction of heterogeneous material systems, each with distinct morphological features and degrees of clustering. Moreover, we showed that the learned representations, in combination with the effective medium theory, can be used to estimate elastic properties of these material systems.




Although the example material systems used here are statistically homogeneous and isotropic two-phase systems, we note that our general procedure can be readily generalized to anisotropic, inhomogeneous and multi-phase material systems. This is because neither the convolution-based encoder nor the gradient-based decoder depends on the isotropy and homogeneity assumptions. However, Bayesian optimization would be challenging as the kernel space for the $n$-point correlations might not be easily characterized by a single parameter distribution. To address this challenge, a possible solution is to formulate the encoder-decoder process as a differential network, which allows direct optimization on the kernel space. In our future work, we will explore this approach, as well as constructing spatial-temporal correlations to represent dynamics of evolving material systems.


\section*{Data availability}
The source codes and data in this work are available \href{https://github.com/shengcheng/Material_Reconstruction_Correlation}{here}.

\section*{Acknowledgement}
This work is supported by the National Science Foundation, Division of Material Research under grant NO. 2020277 (AI Institute: Planning: Novel Neural Architectures for 4D Materials Science)

\bibliography{mybibfile}

\end{document}